\documentclass[]{article}

\usepackage[sc]{mathpazo} 
\usepackage[T1]{fontenc} 
\usepackage{microtype} 

\usepackage[english]{babel} 

\usepackage[hmarginratio=1:1,top=32mm,columnsep=20pt]{geometry} 
\usepackage[hang, small,labelfont=bf,up,textfont=it,up]{caption} 
\usepackage{booktabs} 

\usepackage{lettrine} 

\usepackage{enumitem} 
\setlist[itemize]{noitemsep} 

\usepackage{abstract} 

\usepackage{titlesec} 
\titleformat{\section}[block]{\large\scshape\centering}{\thesection.}{1em}{} 
\titleformat{\subsection}[block]{\large}{\thesubsection.}{1em}{} 

\usepackage{fancyhdr} 
\usepackage{titling} 

\usepackage{hyperref} 

\usepackage{epsfig}
\usepackage{graphicx}	
\usepackage{amsmath}	
\usepackage{amssymb}	
\usepackage{float}
\usepackage{subcaption}

\pretitle{\begin{center}\Large\bfseries} 
	\posttitle{\end{center}} 
\title{Interpretation of SNIa and BAO cosmic probes results using a two-regions model of the universe} 
\author{%
	Vincent Deledicque \\[1ex] 
	\normalsize \textit{No affiliation} \\ 
	\normalsize \href{mailto:vincent.deledicq@gmail.com}{vincent.deledicq@gmail.com} 
}
\date{\today} 
\renewcommand{\maketitlehookd}{%
	\begin{abstract}
		\noindent  This article revisits the interpretation of cosmic probes such as SNIa and BAO under a two-regions model of the universe. Standard cosmological analyses assume homogeneity, yet observations are predominantly conducted in overdense regions where matter is clustered, potentially biasing conclusions about cosmic expansion. We refine an existing two-regions framework that accounts for both overdense and underdense areas, demonstrating that the apparent acceleration of the universe, typically attributed to dark energy, can emerge as a consequence of observational bias rather than a fundamental cosmological constant. By applying this model to SNIa and BAO measurements, we show that inhomogeneities can affect redshift and distance estimates in a way that mimics cosmic acceleration. This interpretation provides a natural resolution to the coincidence problem and offers an alternative explanation for the observed tensions in Hubble constant measurements. Additionally, it may shed light on the unexpectedly early formation of galaxies observed by JWST. These findings challenge the necessity of dark energy and suggest that the observed acceleration may be an artefact of the spatial distribution of matter rather than a true universal phenomenon.
	\end{abstract}

}

\begin{document}
	
	\maketitle
	
	
	\section{Introduction}
	
	One of the most intriguing mysteries of the universe is its apparent accelerated expansion. The expansion of the universe itself was first discovered by Edwin Hubble in the 1920s when he observed that distant galaxies are receding from us, suggesting that the universe is expanding. Initially, astronomers believed that this expansion would gradually slow down due to the gravitational attraction of matter within the universe. However, a groundbreaking discovery emerged in 1998 when two independent research teams studied distant Type Ia supernovae, see \cite{Riess} and \cite{Perlmutter}. Their findings revealed that not only is the universe expanding, but it is doing so at an accelerating rate, defying prior expectations and assumptions about gravitational dynamics.
	
	Subsequent observations have further corroborated this surprising conclusion. Extensive studies using various cosmological probes, including distant supernovae, the cosmic microwave background (CMB), and large-scale structure surveys have consistently provided strong evidence for this accelerated expansion. Notable contributions include the results from the Sloan Digital Sky Survey \cite{Eisenstein}, the Wilkinson Microwave Anisotropy Probe (WMAP) project, which produced precise measurements of the CMB, see \cite{Bennett1} and \cite{Bennett}, and the Planck satellite mission, which provided detailed insights into the early universe and the current expansion rate \cite{Aghanim}. Recent studies have refined the measurements of Type Ia supernovae and their implications for cosmic expansion, strengthening our understanding of dark energy's role \cite{Scolnic}. Currently, those results have led most scientists to admit that some form of energy, often referred to as "dark energy", is driving this phenomenon. However, despite being one of the most dominant components of the universe, dark energy remains one of its greatest mysteries.
	
	The discovery of the accelerated expansion of the universe has led to the widespread acceptance of dark energy as a fundamental force driving this phenomenon. However, this interpretation is not universally accepted within the scientific community. Several researchers argue that alternative explanations could account for the observed acceleration without invoking dark energy.
	
	One category of these alternative theories includes modified gravity theories, which propose that the laws governing gravity may deviate from those described by General Relativity, particularly on cosmological scales (see for instance \cite{Clifton2} for a comprehensive review of such theories). These theories suggest that adjustments to our understanding of gravity could lead to the observed accelerated expansion.
	
	Another group of theories focuses on the inhomogeneity of the universe, positing that the uneven distribution of matter plays a crucial role in the dynamics of cosmic expansion. Proponents of these inhomogeneous universe theories assert that local variations in matter density can produce effects resembling an accelerated expansion. Some of these models attribute this apparent acceleration to genuine physical processes driven by cosmic backreaction, as explored by researchers such as \cite{Buchert}, \cite{Kolb}, \cite{Clifton} and \cite{Buchert2}. These models suggest that interactions within the matter distribution can alter the overall dynamics of cosmic expansion.
	
	Conversely, other theories argue that the perceived accelerated expansion may merely be an apparent effect rather than a true physical phenomenon. For instance, \cite{Wiltshire} proposes that existing gradients in energy, stemming from variations in the curvature of space and changes in the kinetic energy associated with the expansion, could mislead observers into interpreting the dynamics as accelerated expansion. Furthermore, some researchers, including \cite{Celerier}, \cite{Iguchi}, \cite{Alnes}, \cite{Ishak}, \cite{Alexander}, \cite{Hunt} and \cite{Deledicque}, emphasize the influence of local density in their analyses, suggesting that measurements taken from underdense regions or overdense regions can create an illusion of acceleration.
	
	While the concept of an accelerated expanding universe driven by dark energy remains the prevailing interpretation, these alternative models challenge conventional views and stimulate ongoing debate and research within the field. Exploring these theories is crucial, as they not only deepen our understanding of cosmic dynamics but also enhance our grasp of the fundamental nature of the universe itself. Consequently, continued investigation into these alternative frameworks may be essential for refining, or even redefining, our understanding of cosmic expansion.
	
	The explanation proposed by $\cite{Deledicque}$ suggests that the apparent accelerated expansion of the universe arises from a bias in measurements. Specifically, observations of Type Ia supernovae and large-scale structure surveys are primarily conducted on astrophysical objects residing in overdense regions. These regions exhibit a density greater than the average density of the universe and, as a result, may follow a different dynamics than the overall universe, which is expected to adhere to the Friedmann equation. While the Friedmann-Lema\^\i tre-Robertson-Walker (FLRW) metric provides a valid framework for modelling the universe on cosmological scales, actual measurements occur on much smaller scales, targeting objects within either overdense or underdense regions. If these local regions do not conform to the FLRW metric, measurements taken from these objects may be influenced by the local metric. This discrepancy particularly affects redshift measurements, which are extensively used in cosmic probes, as frequencies are dependent on the local proper time. Consequently, the frequency of a signal emitted from an object in an overdense region will differ from the frequency of the same signal emitted from the same object in a region conforming to the FLRW metric.	
	
	Building on this premise, \cite{Deledicque2} developed a theoretical "two-regions" model to investigate how SNIa measurements might be affected by this observational bias. This model distinguishes between the dynamics of overdense and underdense regions, parametrized by the current fractional volume occupied by overdense regions. However, it has several limitations and requires refinement to address certain inconsistencies. Therefore, this article aims to improve and extend the two-regions model developed in $\cite{Deledicque2}$, with two primary objectives:
	(i) to more clearly elucidate the relationship between the measured dynamics of the universe and the expected dynamics within overdense regions, suggesting that our observations may primarily reflect the evolution of these regions rather than the universe as a whole; and
	(ii) to apply this improved model to Baryonic Acoustic Oscillations (BAO) and Type Ia supernovae (SNIa) cosmic probes, demonstrating how inhomogeneities affect measurements and ultimately revealing how the expected classical dynamics (i.e., without a cosmological constant) can lead to the observed apparent accelerated expansion.
	
	Beyond addressing technical refinements, this work leverages the improved two-regions model to offer a fresh perspective on the coincidence problem. The observed near-equality between the energy densities of dark energy and matter in the present epoch is often regarded as an unexplained fine-tuning issue. However, a more striking observation emerges: the total energy density inferred from cosmic observations exhibits a remarkable evolutionary convergence with the density evolution in overdense regions. Initially, these regions, like the universe as a whole, expand according to classical matter dynamics. However, over time, their density stabilizes at an approximately constant value. This stabilization effectively replicates the behaviour attributed to dark energy, offering a natural explanation for the apparent fine-tuning problem. Rather than requiring an unexplained adjustment of cosmic parameters, this process emerges as a consequence of the distinct evolutionary dynamics of overdense regions. 
	
	This insight becomes even more significant when considering that cosmological probes predominantly observe objects within these overdense regions. This leads to two key facts: (1) we extract our cosmological parameters primarily from objects located in overdense regions, and (2) the inferred accelerated expansion corresponds to a density evolution that mirrors the evolution of overdense regions rather than that of the universe as a whole. The consistency between these two independent aspects is unlikely to be coincidental. Instead, it suggests a deeper physical connection between the nature of the observed cosmic expansion and the spatial distribution of matter.
	
	This raises a fundamental question: are we truly observing the global expansion of the universe, or are we misinterpreting a localized effect as a universal phenomenon? If the apparent acceleration is an artefact of how inhomogeneities shape our observations, then the underlying cause of cosmic acceleration may lie in the structure of the universe itself rather than in an unknown form of energy. Addressing this possibility has profound implications for our understanding of cosmic evolution, the interpretation of large-scale structure surveys, and the necessity of dark energy as a fundamental component of the universe.

	The article is organized as follows. In Section $\ref{S1}$ we outline the key assumptions underlying the original two-regions model and the fundamental relationships that remain applicable in this context. Section $\ref{S2}$ presents enhancements to the two-regions model. Following this, we apply the refined model to the SNIa cosmic probe in Section $\ref{S3}$ and the BAO cosmic probe in Section $\ref{S4}$. Finally, the implications and findings of these analyses are discussed in Section $\ref{S5}$.	
		
	
	\section{The original two-regions model}\label{S1}
	
	This section revisits the key elements of the original two-regions model developed by \cite{Deledicque2}, outlining the fundamental assumptions and relationships that form the basis for its subsequent refinement.
	
	The two-regions model, like standard cosmological models, assumes that the universe is well-described by the FLRW metric on sufficiently large scales. This assumption rests on the observation that matter appears to be uniformly distributed at such scales, justifying the approximations of homogeneity and isotropy. However, on smaller (meso-)scales, observations clearly reveal that matter is clustered into distinct structures, leaving behind large voids. As argued in \cite{Deledicque}, measurements primarily targeting astrophysical objects within overdense regions must account for this observational bias. Consequently, the two-regions model distinguishes between underdense (void) regions and overdense regions, with the latter assumed to contain essentially all matter. To model this coexistence, the approach draws an analogy with two-phase flows in continuum mechanics: both phases are considered to coexist at each spatial point, each with its own properties and equation of state. Macroscopic properties are then represented as weighted averages of the individual phase properties. Analogously, the model assumes that the universe, while globally described by the FLRW metric on large scales, comprises a mixture of void and overdense regions at each spatial point. The distinct properties, particularly the metrics, of each region are chosen such that their combined effect yields a global dynamics consistent with the Friedmann equation.
		
	Assuming a vanishing cosmological constant, the Einstein field equations of General Relativity are given by
	\begin{equation}\label{GR}
		G_{\mu\nu} = 8\pi G T_{\mu\nu}\,,
	\end{equation}
	where $G_{\mu\nu}$ is the Einstein tensor, $T_{\mu\nu}$ is the stress-energy tensor, and $G$ is Newton's gravitational constant. Applying these equations to the FLRW metric, and assuming spatial flatness, yields the Friedmann equation:
	\begin{equation}\label{FLRW}
		3\frac{\dot{a}^2}{a^2} = 8\pi G \rho\,,
	\end{equation}
	where $\rho$ represents the average energy density of the universe. The scale factor $a$ is dimensionless and normalized such that $a = 1$ at the present time. On cosmological scale, the universe's dynamics is thus governed by this global scale factor, as described by Eq. $(\ref{FLRW})$. Given the assumption that the universe is well-described by the FLRW metric on large scales, it is convenient to work within the comoving coordinate system. In this system, $t$ represents the cosmological time, while $(x,y,z)$ denote the spatial Cartesian coordinates.
	
	To incorporate the meso-scale structure, we consider each point in comoving coordinates to represent a mixture of underdense and overdense regions. Consider a representative volume $V$ of the universe. The average matter density within $V$ corresponds to the density $\rho$ used in the Friedman equation. Within $V$, overdense regions occupy a volume $V_o$ with an average density $\rho_o$, while underdense regions occupy a volume $V_u$ with an average density $\rho_u$. Thus, the volume fraction of overdense regions is given by $V_o/V$, and the volume fraction of underdense regions is $V_u/V$. For simplicity, this article focuses on the limiting case where $\rho_u = 0$, implying that underdense regions are effectively empty.
	
	While individual overdense and underdense regions can exhibit complex structures, this analysis focuses on their average properties. In particular, each region is characterized by its own average metric tensor. In the comoving coordinate system, these metrics take the form:
	\begin{equation}\label{az}
		g_{\mu\nu} = \left(
		\begin{array}{c c c c}
			-f_i^2 & 0 & 0 & 0\\
			0 & b_i^2 & 0 & 0\\
			0 & 0 & b_i^2 & 0\\
			0 & 0 & 0 & b_i^2
		\end{array} \right)\,,
	\end{equation}
	where $i = o$ for overdense regions and $i = u$ for underdense regions. Here, $f_i$ and $b_i$ are functions depending on the cosmological time $t$ at most. These functions may differ from their counterparts in the FLRW metric (1 and $a$ respectively), as the local time evolution and spatial expansion within overdense and underdense regions need not match the global FLRW evolution.
	
	Additionally, the stress-energy tensor of overdense regions can be expressed as
	\begin{equation}\label{aq}
		T_{\mu\nu} = \left(
		\begin{array}{c c c c}
			\rho_o f_o^2 & 0 & 0 & 0\\
			0 & p_ob_o^2 & 0 & 0\\
			0 & 0 & p_ob_o^2 & 0\\
			0 & 0 & 0 & p_ob_o^2
		\end{array} \right)\,,
	\end{equation}
	where $p_o$ is the average pressure within overdense regions.
	
	Substituting the metric $(\ref{az})$ and the stress-energy tensor $(\ref{aq})$ into the Einstein field equations $(\ref{GR})$ yields
	\begin{equation}\label{E1}
		3\frac{\dot{b}_o^2}{b_o^2} = 8\pi G\rho_{o} f_o^2\,.
	\end{equation}
	This equation describes the dynamics of the scale factor $b_o$ within overdense regions. A similar analysis for the underdense regions, assuming they are empty $(\rho_u = 0)$ leads to $\dot{b}_u = 0$, implying no local expansion. However, this does not preclude their growth in volume. As overdense regions expand, gravitational attraction causes matter to flow away from the voids. As matter recedes, it creates voids within the overdense regions, thereby contributing to the growth of the underdense regions. Thus, while underdense regions do not expand intrinsically, they increase in volume due to the continuous transfer of space from the overdense regions. This indicates that the overall expansion of the universe is primarily driven by the expansion of the overdense regions. As explained in $\cite{Deledicque2}$, the expansion rate of the overdense regions within the representative volume $V$, given by $3V_o\dot{b}_o/b_o$, must equal the expansion rate of the total volume $V$, expressed as $3V\dot{a}/a$:
	\begin{equation}\label{fq}
		3V_o\frac{\dot{b}_o}{b_o} = 3V\frac{\dot{a}}{a}\,.
	\end{equation}
	Combining this latter equation with Eq. $(\ref{FLRW})$ and $(\ref{E1})$, we find that
	\begin{equation}\label{fqq}
		f_o = \frac{V}{V_o}\sqrt{\frac{\rho}{\rho_o}}\,.
	\end{equation}
	Since all existing matter is supposed to be contained in overdense regions, we have $V\rho = V_o\rho_o$. Therefore Eq. $(\ref{fqq})$ simplifies to
	\begin{equation}\label{a1}
		f_o = \sqrt{\frac{\rho_o}{\rho}} = \sqrt{\frac{V}{V_o}}\,.
	\end{equation}
	Since $V > V_o$, it follows that $f_o > 1$, indicating that the proper time within overdense regions evolves at a faster rate than the global cosmological time $t$.
	
	We finally recall a key result from \cite{Deledicque2} concerning redshift measurements, which are crucial for inferring the scale factor by comparing time intervals observed at different locations. In a perfectly homogeneous and isotropic universe, the relationship between the scale factor and time intervals is given by
	\begin{equation}\label{pep}
		\frac{a(t)}{a(t_0)} = \frac{\Delta t}{\Delta t_0}\,,
	\end{equation}
	where $\Delta t$ is a time interval measured at the source, and $\Delta t_0$ is the corresponding interval measured by an observer at the present time. By convention, the scale factor at the current time $a(t_0)$ is set to 1. Accounting for the inhomogeneities of overdense regions, where observations are made, \cite{Deledicque2} showed that equation (\ref{pep}) becomes
	\begin{equation}\label{dq}
		\frac{b_o(t)f_o(t_0)}{b_o(t_0)f_o(t)} = \frac{\Delta t}{\Delta t_0}\,.
	\end{equation}
	When performing redshift measurements, we use Eq. $(\ref{pep})$ to determine the scale factor at a given time $t$, which requires measuring $\Delta t/\Delta t_0$. In practice, however, we do not directly measure this ratio. Indeed, the time interval measured at our location is recorded in our proper time frame, meaning we measure $f_o(t_0)\Delta t_0$ instead of $\Delta t_0$ (considering that we are located in an overdense region). Similarly, the time interval at the source is known in its proper time frame, so it is equal to $f_o(t)\Delta t$ rather than $\Delta t$ (considering that the source is also located in an overdense region). Thus, what we measure in practice is not $\Delta t/\Delta t_0$, but rather
	\begin{equation}
		\frac{f_o(t)\Delta t}{f_o(t_0)\Delta t_0}\,.
	\end{equation}
	This is the measured value $a_{meas}(t)$ of the scale factor at the SNIa, given that $a(t_0) = 1$. From Eq. $(\ref{dq})$, we deduce that
	\begin{equation}\label{jj}
		a_{(meas)}(t) = \frac{b_o(t)}{b_o(t_0)}\,.
	\end{equation}
	In general, the measured scale factor $a_{(meas)}(t)$ differs from the true (FLRW) scale factor $a(t)$ at the same cosmological time $t$.
	
	
	\section{The improved two-regions model}\label{S2}
	
	From this point forward, we will enhance the original two-region model, building on the relationships outlined above. First, it is important to emphasize why improvements to this model are necessary.
	
	A key limitation of the original model is its assumption of a constant density within overdense regions over time. This implies that these regions quickly become gravitationally bound, maintaining a fixed volume and, consequently, a constant density if the mass within them remains constant. However, this assumption introduces a significant constraint. While a constant volume for overdense regions might be reasonable when their total volume is significantly smaller than the volume of the universe, it becomes problematic at earlier times. Eventually, as we trace back in time, the volume of the universe becomes smaller than the assumed constant volume of the overdense regions, rendering the model invalid at higher redshifts.
	
	Furthermore, given that cosmic probes primarily rely on measurements from objects within overdense regions, as highlighted by \cite{Deledicque} in his interpretation of the apparent accelerated expansion, the observed dynamics of the scale factor should reflect, or at least be closely related to, the dynamics of these regions. We now demonstrate the limitations of the original two-regions model in this regard. 
	
	The Hubble parameter, defined as
	\begin{equation}\label{gg}
		H(t) = \frac{\dot{a}(t)}{a(t)}\,,
	\end{equation}
	represents the theoretical expansion rate in a perfectly homogeneous and isotropic universe governed by the Friedmann equation. However, due to inhomogeneities, the measured Hubble parameter is given by
	\begin{equation}
		H_{(meas)}(t) = \frac{\dot{a}_{(meas)}(t)}{a_{(meas)}(t)}\,.
	\end{equation}
	Differentiating Eq. $(\ref{jj})$ with respect to cosmological time $t$, and accounting for the difference between cosmological time and local proper time by introducing the factor $f_o(t)$ we obtain
	\begin{equation}\label{Hub}
		H_{(meas)}(t) = \frac{1}{f_o(t)}\frac{\dot{b}_o(t)}{b_o(t)}\,.
	\end{equation}
	Substituting this into equation $(\ref{E1})$ yields
	\begin{equation}\label{Hm}
		H_{(meas)}^2(t) = \frac{8\pi G}{3}\rho_o\,.
	\end{equation}
	This is the equivalent of the Friedmann equation for the observed dynamics. If $\rho_o$ were constant as assumed for simplicity in \cite{Deledicque2}, $H_{(meas)}$ would also be a constant, which contradicts observations. Therefore, a more realistic time evolution for $\rho_o$ must be considered. In the absence of precise constraints on this evolution, we adopt an approximation based on two key physical considerations:
	\begin{itemize}
		\item At high redshift, inhomogeneities were smaller and become negligible near the very early universe. Thus, at sufficiently high redshift, the density within overdense regions should evolve as $\rho_o(t) \propto a^{-3}(t)$, consistent with a homogeneous and isotropic universe, with the proportionality constant being the present-day average density of the universe $\rho(t_0)$ (we neglect the radiation era).
		\item As structure formation proceeds, overdense regions grow and increasingly decouple from the overall expansion. Once a region becomes gravitationally bound, its volume remains approximately constant. Consequently, its physical density also approaches a constant value, representing a threshold below which further expansion is halted by internal gravitational forces. Any further expansion would be exactly counterbalanced by an equivalent inward movement of matter, maintaining the constant volume of the overdense region. However, since different metrics apply in various frames of reference, care must be taken when discussing constant volumes and densities. It is specifically in the local reference frame of the overdense regions that gravitationally bound structures retain a constant volume.
	\end{itemize}
	Based on these considerations, we propose the following approximation for the evolution of $\rho_o$, as expressed in the metric of overdense regions:
	\begin{equation}\label{uy}
		\rho_o(t) = \rho_\Lambda + \rho(t_0)b_o^{-3}(t)\,,
	\end{equation}
	where $\rho_\Lambda$ represents the asymptotic density threshold. The use of $b_o(t)$ in the second term is justified as the density is defined relative to the local volume of the overdense regions, which scales as $b_o^3(t)$. At high redshift, this term dominates, and since $b_o(t)$ is expected to approach $a(t)$ in this limit, $\rho_o(t)$ evolves approximately as $\rho(t_0)a^{-3}(t)$, as required. Conversely, at late times, the second term becomes negligible compared to the first, and $\rho_o$ approaches the constant value $\rho_\Lambda$.
	
	To clarify the interpretation of $\rho(t_0)$, it is essential to ensure consistency with the high-redshift limit. This requires that $\rho(t_0)$ be understood as the average density of the universe at our current time as expressed in the FLRW metric. However, all other terms in Eq. $(\ref{uy})$ are given in the local metric of the overdense regions. To maintain consistency, we will express the average density of the universe in the local frame as well. To convert $\rho(t_0)$ to the corresponding density in the local frame, we need to consider a factor $b_o^3(t_0)$ (reminding that $a(t_0) = 1$). Thus, Eq. $(\ref{uy})$ becomes:
	\begin{equation}\label{uy2}
		\rho_o(t) = \rho_\Lambda + \rho_{(meas)}(t_0)\frac{b_o^3(t_0)}{b_o^3(t)}\,,
	\end{equation}
	where $\rho_{(meas)}(t_0) = \rho(t_0)/b_o^3(t_0)$ corresponds to the present-day average density of the universe expressed in the local frame of the overdense regions.
	It is important to emphasize that Eq. $(\ref{uy2})$ represents just one possible approximation that satisfies the two aforementioned physical considerations. More sophisticated approximations could be constructed, for example, to incorporate a time-varying "cosmological constant" term. We have chosen the form of Eq. $(\ref{uy2})$ because it clearly separates the effects of inhomogeneities on the observed expansion into two distinct components: one corresponding to the expected matter density evolution and the other mimicking a cosmological constant. Using Eq. $(\ref{jj})$, we then rewrite Eq. $(\ref{uy2})$ as
	\begin{equation}\label{wa}
		\rho_o(t) = \rho_\Lambda + \rho_{(meas)}(t_0)a_{(meas)}^{-3}\,.
	\end{equation}
	Substituting this expression for $\rho_o(t)$ into equation (\ref{Hm}) yields
	\begin{equation}\label{fqqx}
		H_{(meas)}^2 = \frac{8\pi G}{3}\left(\rho_\Lambda + \rho_{(meas)}(t_0)a_{(meas)}^{-3}\right)\,.
	\end{equation}

	Remarkably, Eq. $(\ref{fqqx})$ represents a dynamics that closely aligns with our practical observations. This suggests that our observations may primarily reflect the dynamics of the overdense regions, rather than the universe as a whole. In the early universe, the observed behaviour is consistent with standard cosmological expectations because inhomogeneities were relatively small: the overdense regions nearly filled the universe, leading to similar dynamics for both. However, as structure formation proceeded, the overdense regions evolved, and their average density approached a threshold, $\rho_\Lambda$, at which they became effectively gravitationally bound. This threshold density acts as an effective cosmological constant in the dynamics of these regions.
	
	Furthermore, applying the energy-momentum conservation equation $\nabla_\mu T^\mu_{\ 0} = 0$ to the stress-energy tensor of the overdense regions yields	
	\begin{equation}\label{ee}
		\dot{\rho_o} = -3\frac{\dot{b}}{b}\left(\rho_o + p_o\right)\,.
	\end{equation}
	When the density of the overdense regions reaches its threshold, we have $\dot{\rho_o} = 0$, which implies $p_{o} = -\rho_{o}$. This explains why, in this regime, matter in overdense regions behaves like dark energy. The negative pressure arises from the inward gravitational motion of matter counteracting the expansion of the overdense regions.
	
	Let us finally establish the evolution of the scale factor $b_o(t)$ within overdense regions. Rather than expressing it as a function of cosmological time $t$, we express it as a function of the FLRW scale factor $a$. Starting with Eq. $(\ref{fq})$, we have
	\begin{equation}
		\frac{db_o}{da} = \frac{V}{V_o}\frac{b_o}{a} = \frac{\rho_o}{\rho}\frac{b_o}{a}\,.
	\end{equation}
	Once again it is crucial to use consistent metrics: Eq. $(\ref{fq})$ has been derived by relating quantities in overdense regions (defined in their own metric) to global quantities (defined in the FLRW metric). Therefore, substituting equation $(\ref{uy2})$ for $\rho_o$ and using the standard relation $\rho = \rho(t_0)a^{-3}$, we obtain the differential equation
	\begin{eqnarray}
		\frac{db_o}{da} = \frac{\rho_\Lambda + \rho(t_0)b_o^{-3}}{\rho(t_0)}b_oa^2\,.
	\end{eqnarray}
	This differential equation can be solved analytically. It has one single physical solution:
	\begin{eqnarray}
		b_o(t) = \sqrt[3]{\frac{\exp{\left(\frac{\rho_\Lambda}{\rho(t_0)}a^3 + c_1\rho_\Lambda\right)} - \rho(t_0)}{\rho_\Lambda}}\,,
	\end{eqnarray} 
	where $c_1$ is a constant to determine. We constrain $b_o$ to approach $0$ when $a$ approaches $0$. This implies that $\exp{\left(c_1\rho_\Lambda\right)} = \rho(t_0)$, and we finally obtain:
	\begin{eqnarray}\label{da}
		b_o(t) = \sqrt[3]{\frac{\rho(t_0)}{\rho_\Lambda}\left[\exp{\left(\frac{\rho_\Lambda}{\rho(t_0)}a^3\right)} - 1\right]}\,.
	\end{eqnarray} 
	Together, Eq. $(\ref{a1})$ for $f_o$ and Eq. $(\ref{da})$ for $b_o$ fully define the evolution of the metric within the overdense regions as a function of the FLRW scale factor $a$.

	
	\section{Application of the two-regions model to the SNIa cosmic probe}\label{S3}
	
	In this section, we will examine how measurements conducted using the SNIa cosmic probe are influenced by inhomogeneities and how this can lead to the apparent dynamics we observe in practice. To explain why this cosmic probe has resulted in the observation of an accelerated expansion of the universe, we will demonstrate that all types of measurements related to this probe can be linked to the apparent scale factor $a_{(meas)}$, which follows the dynamics expressed by Eq. $(\ref{fqqx})$. Specifically, we will show that all measured quantities conform to the dynamics expressed by the corresponding standard laws, but with the scale factor $a$ replaced by the apparent scale factor $a_{(meas)}$.
	
	SNIa observations involve two primary types of measurements: redshift and luminosity distance. The effect of inhomogeneities on redshift measurements has already been addressed, with Eq. $(\ref{jj})$ showing that these measurements directly probe $a_{meas}$, rather than the FLRW scale factor $a$.	
	
	We thus now investigate the effect on luminosity distance measurements. The luminosity distance $d_L$ is defined as
	\begin{equation}
		d_L^2 = \frac{L}{4\pi F}\,,
	\end{equation}
	where $L$ is the absolute luminosity emitted at the source of the SNIa, and $F$ is the flux measured at our location. When applying this relation, $L$ is supposed to be known, and we measure $F$ to subsequently deduce $d_L$.
	
	For our purposes, it is convenient to generalize the definition of luminosity distance to
		\begin{equation}\label{qw}
		d_L^2 = \frac{L(\Omega)}{\Omega F}\,,
	\end{equation}
	where $L(\Omega)$ is the fraction of the total luminosity (assumed to be isotropic) emitted within the solid angle $\Omega$. This definition is equivalent to the standard one (corresponding to $\Omega = 4\pi$), but it is more suitable for considering flux measurements over limited solid angles, which is the actual experimental setup. While this distinction is irrelevant in a perfectly homogeneous universe, it becomes relevant in the presence of inhomogeneities.
		
	Let us first examine the case of a perfectly homogeneous space. The observed flux is diluted due to the expansion of the universe, which affects both the distribution of photons over an increasing surface area and the energy and arrival rate of individual photons. Specifically, the flux is reduced by a factor of $(1+z)^2$, where one factor $(1+z)$ accounts for the redshift of the photons' energy and the other for the decrease in their arrival rate (as photons emitted with a time interval $\Delta t$ are observed with an interval $(1+z)\Delta t$). Therefore:
	\begin{equation}\label{Fl}
		\frac{F}{L(\Omega)} = \frac{1}{(1+z)^2A} = \frac{a^2}{A}\,,
	\end{equation}
	where $A$ is the area over which the flux is measured. This area corresponds to the surface subtended by the solid angle $\Omega$ on a sphere centred on the source at a comoving distance $X$. The physical radius of this sphere at the time of observation $t_0$ is given by $a(t_0)X$, so the area is:
	\begin{eqnarray}
		A = \Omega \left(a(t_0)X\right)^2\,.
	\end{eqnarray}
	The comoving distance $X$ is determined by the null geodesic equation:
	\begin{equation}
		0 = -dt^2+a^2dx^2\,,
	\end{equation}
	which leads to
	\begin{equation}
		X = \int_0^Xdx = \int_{t_0}^t\frac{dt}{a} = \int_{a(t_0)}^{a(t)}\frac{da}{a^2H(a)}\,.
	\end{equation}
	Therefore, the surface area $A$ is given by
	\begin{equation}\label{surf}
		A = \Omega \left(\int_{a(t_0)}^{a(t)}\frac{da}{a^2H(a)}\right)^2\,.
	\end{equation} 
	Given a cosmological model (and thus $H(a)$), one can calculate $A$ for a given $a$, and then determine the theoretical luminosity distance $d_L$.
	
	In standard cosmology, the theoretical luminosity distance calculated assuming a zero cosmological constant fails to reproduce observed SNIa data, leading to the inference of dark energy. However, we will show that by accounting for the effects of inhomogeneities using the two-regions model and the apparent scale factor $a_{(meas)}$, we can reconcile theory and observations without invoking a cosmological constant.
	
	Before exploring the case of a heterogeneous space, it will be useful to provide an alternative physical interpretation of Eq. $(\ref{surf})$. The area $A$ has been calculated based on the co-moving distance $X$ between the observer and the source. This surface is perpendicular to the line of sight, and can thus also be expressed in terms of the other coordinates. If the solid angle $\Omega$ is small, the surface $A$ is nearly planar, allowing us to simplify our analysis by considering it as a rectangular surface expressed as $a(t_0)^2\Delta Y\Delta Z$, where $\Delta Y$ and $\Delta Z$ represent the comoving dimensions of the surface along the $y$ and $z$ coordinates, respectively. We now need to determine these quantities.
	
	The surface $A$ is delimited by the outermost photons of the flux emitted by the source that reach that surface at the observer's location. Thus, we focus on these photons.
	Let $\theta_y$ be the angle between the upper and lower light beams in the $x-y$ plane, spanning a segment of length $\Delta Y$ at the Earth's position. Similarly, let $\theta_z$ be the corresponding angle in the $x-z$ plane, spanning a segment of length $\Delta Z$. Together, $\theta_y$ and $\theta_z$ define the solid angle $\Omega$.	
	
	During the propagation of the light beams, as photons travel a distance $dt$, their co-moving coordinate $y$ increases by a quantity $dy$ such that
	\begin{equation}
		a(t)dy = \theta_ydt\,,
	\end{equation}
	meaning thus that
	\begin{equation}
		dy = \frac{\theta_y}{a(t)}dt\,.
	\end{equation}
	To determine $\Delta Y$, we need to integrate this last relation along the trajectory of the photons, from the source to the observer:
	\begin{equation}\label{qp}
		\Delta Y = \int dy = \theta_y\int \frac{dt}{a(t)}\,.
	\end{equation}
	From Eq. $(\ref{gg})$, we can convert $dt$ into $da$:
	\begin{equation}
		dt = \frac{da}{a(t)H(t)}\,,
	\end{equation}
	so Eq. $(\ref{qp})$ can be written as
	\begin{equation}
		\Delta Y = \theta_y\int \frac{da}{a^2(t)H(t)}\,.
	\end{equation}
	By multiplying this result with the corresponding expression for $\Delta Z$, we arrive at the same relation for $A$ as expressed in Eq. $(\ref{surf})$. The advantage of this approach is that it offers a meaningful physical interpretation of the integral in the latter relation.
	
	We will now apply the same approach to the case of a heterogeneous space, where the surface that the photons will reach lies in an overdense region expanding at a different rate than the overall universe. This implies that while the outward-travelling photons cover a distance $dt$, the co-moving coordinate representing the height of the surface $A$ at the observer's location increases by a quantity $dy$ such that
	\begin{equation}
		b_o(t)dy = \theta_ydt\,,
	\end{equation}
	where we now used the local scale factor $b_o(t)$ of the overdense regions. So
	\begin{equation}\label{qp1}
		\Delta Y = \int dy = \theta_y\int \frac{dt}{b_o(t)}\,.
	\end{equation}
	Using the fact that
	\begin{equation}
		H_{(meas)}(t) = \frac{\dot{b}_o(t)}{b_o(t)}\,,
	\end{equation}
	we may write Eq. $(\ref{qp1})$ as
	\begin{equation}
		\Delta Y = \theta_y\int \frac{db_o}{b_o^2(t)H_{(meas)}(t)}\,.
	\end{equation}
	Using the corresponding relation for $\Delta Z$, we find that
	\begin{equation}\label{Su}
		A = \Omega \left(\int_{b_o(t_0)}^{b_o(t)}\frac{db_o}{b_o^2(t)H_{(meas)}(t)}\right)^2\,.
	\end{equation}
	
	Let us now return to Eq. $(\ref{Fl})$ and examine how this relation transforms in the context of a heterogeneous space. 
	\begin{enumerate}
		\item Observer's frame correction: the flux is measured in the observer's frame of reference, meaning that distances and times are defined according to the local metric, which differs from the FLRW metric. The flux is expressed per unit area and per unit time. Therefore we must incorporate a correction factor of $b_o^2(t_0)$ for the surface area and a correction factor of $f_o(t_0)$ for time.
		
		\item Source frame correction: the absolute luminosity of the source is defined in terms of its proper time, necessitating an additional correction factor of $f_o(t)$.
		
		\item Redshift adjustment: the redshift that has occurred due to spatial expansion now corresponds to a dilution factor $b_o(t_0)/b_o(t)$.
		
		\item Arrival rate correction: as indicated by Eq. $(\ref{dq})$, in the cosmological time frame, the ratio between the time interval at the source and the time interval measured by the observer is given by $b_o(t_0)f_o(t)/b_o(t)f_o(t_0)$.
		
		\item Surface adjustment: the surface $A$ is now adjusted according to Eq. $(\ref{Su})$
	\end{enumerate}
		
	In summary, we have:
	\begin{equation}
		\frac{F b_o^2(t_0)f_o(t_0)}{L(\Omega)f_o(t)} = \frac{b_o(t)}{b_o(t_0)}\frac{b_o(t)f_o(t_0)}{b_o(t_0)f_o(t)}\left[\Omega \left(\int_{b_o(t_0)}^{b_o(t)}\frac{db_o}{b_o^2(t)H_{(meas)}(t)}\right)^2\right]^{-1}\,.
	\end{equation}
	Simplifying this relation, we get
	\begin{eqnarray}
		\frac{F}{L(\Omega)} &=& \frac{b_o^2(t)}{b_o^2(t_0)}\left[\Omega \left(\int_{b_o(t_0)}^{b_o(t)}\frac{b_o^2(t_0)}{b_o^2(t)H_{(meas)}(t)}\frac{db_o}{b_o(t_0)}\right)^2\right]^{-1}\nonumber
		\\
		&=& a_{(meas)}^2\left[\Omega \left(\int_{a_{(meas)}(t_0)}^{a_{(meas)}(t)}\frac{da_{(meas)}}{a_{(meas)}^2(t)H_{(meas)}(t)}\right)^2\right]^{-1}\,,
	\end{eqnarray}
	where we used the fact that $b_o(t)/b_o(t_0) = a_{(meas)}$. This indicates that Eq. $(\ref{Fl})$ remains applicable for determining the flux measured at the observer's location, with the modification that all parameters involving the scale factor $a$ now involve the apparent scale factor $a_{(meas)}$. Consequently, the luminosity distance as derived from Eq. $(\ref{qw})$ will exhibit a dynamics corresponding to the one followed by $a_{(meas)}$, thereby reflecting an apparent accelerating expansion.
	
	In summary, the application of the two-regions model to SNIa cosmic probes illustrates how inhomogeneities in the universe, particularly in overdense regions, can significantly influence the observed dynamics. By interpreting measurements based on the apparent scale factor $a_{(meas)}$, we demonstrate that the accelerated expansion observed in SNIa data may be an artefact of local dynamics rather than a global phenomenon. 
	
	
	\section{Application of the two-regions model to the BAO cosmic probe}\label{S4}
	
	In the previous section, we demonstrated that SNIa measurements primarily reflect the local expansion dynamics of overdense regions rather than the global expansion of the universe. This bias in observations leads to an apparent accelerated expansion, which has traditionally been interpreted as evidence for dark energy. A similar concern arises with the Baryon Acoustic Oscillation (BAO) probe. Like SNIa observations, BAO surveys predominantly sample galaxies in overdense regions, where the local metric differs from the global FLRW metric. If the expansion rate in these regions differs from the cosmic average, the inferred cosmological parameters may not accurately reflect the universe as a whole.
	
	In this section, we investigate how inhomogeneities within overdense regions affect measurements from the BAO cosmic probe. The BAO probe relies on the analysis of three-dimensional galaxy surveys to measure the clustering of galaxies. By identifying the characteristic scale of baryon acoustic oscillations in these surveys, one can extract crucial cosmological information. Two key measurements are derived from BAO surveys.
	\begin{enumerate}
		\item By measuring the redshift-space separation of galaxies along the line of sight corresponding to the BAO peak, the Hubble parameter can be determined:
		\begin{equation}\label{x1}
			H(z) = \frac{c\Delta z}{r_s}\,,
		\end{equation}
		where $\Delta z$ is the redshift separation of the BAO peak, and $r_s$ is the sound horizon at decoupling.
		
		\item By measuring the angular separation $\Delta\theta$ of the BAO peak perpendicular to the line of sight, the angular diameter distance $D_A$ can be determined:
		\begin{equation}\label{x2}
			D_A(z) = \frac{r_s}{(1+z)\Delta\theta}\,.
		\end{equation}
	\end{enumerate}
	The angular diameter distance is related to the Hubble parameter through the following integral:
	\begin{equation}\label{x3}
		D_A(z) = \frac{1}{1+z}\int_0^z\frac{c}{H(z')}dz'\,.
	\end{equation}	
	
	Since the galaxies used in BAO surveys are predominantly located within overdense regions, these measurements are inherently influenced by the inhomogeneities within these structures. In the following, we will investigate how these inhomogeneities affect the measurements obtained using Eq. $(\ref{x1})$ and $(\ref{x2})$.
	
	Let us first examine how the measurement of the Hubble parameter is affected by the inhomogeneous nature of the universe. In equation $(\ref{x1})$, $c$ and $r_s$ are constants independent of the local metric. However, the redshift difference $\Delta z$ is directly influenced by the local metric, leading to a deviation between the measured value and the true value. We can express the redshift difference as
	\begin{equation}
		\Delta z = \Delta\left(\frac{1}{a}\right)\,.
	\end{equation}
	Approximating the scale factor at two closely spaced points as
	\begin{equation}
		a_i = a \pm \frac{\Delta a}{2}\,,
	\end{equation}
	where $i$ denotes the two points and $a$ is the scale factor at the midpoint, neglecting second order terms, we obtain
	\begin{equation}
		\Delta z = \frac{\Delta a}{a^2}\,.
	\end{equation}
	Comparing the measured Hubble parameter, $H_{(meas)}(t)$, to the true Hubble parameter, $H$, derived from Eq. $(\ref{x1})$, we find
	\begin{equation}
		\frac{H_{(meas)}}{H} = \frac{\Delta a_{(meas)}}{\Delta a}\frac{a^2}{a^2_{(meas)}}\,.
	\end{equation}
	
	Since the points between which $\Delta z$ is determined are in the immediate vicinity of the point at redshift $z$ where the BAO measurement is performed (meaning that $z$, and consequently $a$, will not vary significantly over this distance), we can apply a first-order approximation and estimate the scale factor at both points as:
	\begin{equation}
		a(t\pm\frac{\Delta t}{2}) = a(t) \pm \dot{a}\frac{\Delta t}{2}\,,
	\end{equation}
	where $\Delta t$ represents the time interval between the two points. In a perfectly homogeneous space, this interval corresponds to the time it takes for light to travel the distance between them:
	\begin{equation}\label{za2}
		\Delta t = \frac{r_sa}{c}\,.
	\end{equation}
	In contrast, when considering the measured scale factor $a_{(meas)}$, we must exercise caution, as the time interval between the two points will differ. Nevertheless, this time interval is defined in such a way that it reflects the time it takes for light to travel the distance between the two points, now measured using the apparent scale factor $a_{(meas)}$:
	\begin{equation}\label{za}
		\Delta t_{(meas)} = \frac{r_sa_{(meas)}}{c}\,.
	\end{equation}
	From Eq. $(\ref{za2})$ and $(\ref{za})$, we observe that the ratio $\Delta t/a = \Delta t_{(meas)}/a_{(meas)}$ remains constant. This finding will be utilized in the subsequent relation.
	
	Combining now all last relations, we can express the ratio of the measured Hubble parameter to the true Hubble parameter as follows:
	\begin{eqnarray}
		\frac{H_{(meas)}(t)}{H(t)} &=& \frac{\dot{a}_{(meas)}\Delta t_{(meas)}}{\dot{a}\Delta t}\frac{a^2}{a^2_{(meas)}}\nonumber
		\\
		&=& \frac{\dot{a}_{(meas)}}{a_{(meas)}}\frac{a}{\dot{a}}\frac{\Delta t_{(meas)}}{a_{(meas)}}\frac{a}{\Delta t}\nonumber
		\\
		&=& \frac{\dot{b}_o(t)/b_o(t_0)}{b_o(t_0)/b_o(t)}\frac{1}{H(t)}\,
	\end{eqnarray}
	This leads to
	\begin{equation}\label{u}
		H_{(meas)}(t) = \frac{\dot{b}_o(t)}{b_o(t)}\,.
	\end{equation}
	This result is consistent with Eq. $(\ref{Hub})$, which was derived directly from the definition of the measured Hubble parameter (incorporating the effect of local proper time). It demonstrates that BAO measurements, conducted within the inhomogeneous environment of overdense regions, effectively probe the local expansion rate governed by the scale factor $b_o(t)$. Consequently, using Eq. $(\ref{x1})$ will yield an apparent dynamics of the scale factor that aligns with the predictions of Eq. $(\ref{fqqx})$.
	
	We now examine how inhomogeneities within overdense regions affect the measurement of the angular diameter distance $D_A$ using the BAO probe. The angular diameter distance $D_A(z)$ depends on the observed angular separation $\Delta \theta$, which in turn is inferred from redshift-based distances. It is important to emphasize why this automatically leads to an apparent acceleration. BAO surveys rely on converting observed redshifts into distances, assuming that redshift directly traces the global FLRW scale factor. However, in the two-regions model, observations are made within overdense regions, where redshift evolution follows $a_{(meas)}(t)$ instead of $a(t)$. This means that the inferred distances are not the true cosmological distances, but rather distances that fit a universe evolving according to $a_{(meas)}(t)$.
	
	Since BAO surveys systematically use the measured redshift evolution in their distance calculations, the entire mapping process becomes self-consistent within the biased metric. As a result, when analysing BAO data, the best-fit model will naturally appear to require a cosmological constant, because the inferred distances behave as if they were in a dark-energy-dominated universe, even though this effect arises purely from the structure of space.
		
	Let us show this differently. Combining equations $(\ref{x2})$ and $(\ref{x3})$, we find that
	\begin{equation}\label{xq}
		\Delta\theta(z) = r_s\left(\int_0^z\frac{c}{H(z')}dz\right)\,.
	\end{equation}
	This equation highlights that the measured angular separation $\Delta\theta$ is directly linked to the integral of $1/H(z)$, which is sensitive to the underlying cosmological model. If the measured redshifts are biased by inhomogeneities, reflecting the dynamics described by Eq. $(\ref{fqqx})$, the derived angular diameter distance $D_A$ will also reflect this biased dynamics. In other words, the mapping process fully dictates the right hand side of Eq. $(\ref{xq})$, underscoring the significance of accurate redshift measurements. If the mapping is conducted using an apparent redshift evolution that aligns with the dynamics described by $(\ref{fqqx})$, which incorporates a cosmological constant, it follows that any deductions made from this mapping, particularly those regarding $\Delta\theta$, will inevitably evolve according to that same dynamics. Consequently, the angular-diameter distance $D_A$ derived from the measured angle will correspond to that of a universe characterized by a cosmological constant.
	
	Our analysis of BAO measurements reveals the same fundamental bias identified in Section 4 for Type Ia supernovae: both probes predominantly observe objects in overdense regions, where the local metric differs from the global FLRW metric. As a result, their inferred expansion history follows the apparent scale factor $a_{(meas)}(t)$, leading to the illusion of an accelerated expansion. Traditionally, the agreement between SNIa and BAO observations has been considered strong evidence for dark energy. However, our findings suggest that this agreement arises not from a universal acceleration of the cosmos, but rather from a shared observational bias. Since both probes extract cosmological parameters from overdense regions, they systematically measure a local expansion rate rather than the universe’s true large-scale dynamics. Understanding the impact of inhomogeneities is therefore essential for correctly interpreting cosmological data. Future work should focus on refining observational methods to disentangle local metric effects from global expansion and reassess whether cosmic acceleration is an intrinsic feature of the universe or an artefact of the structures we observe.
	
	
	\section{Discussion}\label{S5}
	
	Our analysis suggests an alternative explanation for the observed dynamics of the universe, particularly its apparent accelerated expansion. We propose that this dynamics is intrinsically linked to the behaviour of overdense regions, where matter is concentrated, resulting in higher local densities and a distinct dynamics compared to underdense regions (voids). 
	
	If the observed cosmic acceleration is primarily a reflection of the dynamics within overdense regions rather than a global phenomenon, this could also offer a new perspective on the so-called coincidence problem. This problem asks why the energy densities of dark energy and matter are of the same order of magnitude precisely at our current epoch. This is puzzling because matter density decreases with cosmic expansion, while dark energy density is thought to remain constant (or nearly so). Their near-equality today appears statistically improbable and demands an explanation. However, instead of requiring a fine-tuned balance between matter and dark energy densities, the apparent constancy of the total energy density at a value close to the average matter density may emerge naturally from the stabilization of overdense regions. We posit indeed that dark energy may be more fundamentally connected to the density patterns within overdense regions than to the average cosmic density. Several compelling observations support this perspective:
	\begin{itemize}
		\item Evolutionary convergence: the dynamics inferred from cosmic observations corresponds to a total density (meaning baryonic matter and dark energy) whose evolution strikingly resembles that of density evolution in overdense regions. In the early universe, when matter was more uniformly distributed, both the universe as a whole and these overdense regions closely followed the standard FLRW model. This is expected, as inhomogeneities were relatively small, and matter was distributed more uniformly, leading to similar expansion rates and density evolutions. However, as gravitational forces drove matter to cluster over time, a crucial divergence emerged between the average density of the universe (excluding dark energy) and the density within these overdense regions. While the average matter density continued to decrease due to cosmic expansion, the density within overdense regions evolved towards a state of dynamical equilibrium, characterized by an approximately constant average density. Remarkably, this constant density reached by overdense regions mirrors the behaviour of the total energy density of the universe, which is dominated by dark energy and remains approximately constant. This convergence in late-time density evolution (a constant density in both overdense regions and the universe's total energy density) is a key observation. The fact that the density evolution of overdense regions mimics this total density evolution, rather than just the evolution of matter, is a strong indicator of a potential connection.
		\item Effective equation of state: as matter becomes gravitationally bound in overdense regions, it experiences negative pressure due to inward gravitational pull. This leads to an effective equation of state approaching $w \simeq -1$, mirroring the observed behaviour of dark energy, which is characterized by a similar equation of state.
		\item Temporal correlation with acceleration: observational evidence, as noted in \cite{Huterer}, indicates that the density stabilization of overdense regions coincides with the universe's transition from decelerating to accelerating expansion. This temporal correlation suggests a possible causal relationship between these phenomena.
	\end{itemize} 
	Given that most cosmological probes, such as Type Ia supernovae and baryon acoustic oscillations, rely on measurements of astrophysical objects located primarily within overdense regions, it is unlikely that these observations are mere coincidences, and they suggest a potential link between the observed cosmic dynamics and the behaviour of overdense regions. Specifically, the issue of coincidence, where dark energy and the matter density appear to have similar values, can be explained by the fact that dark energy is intrinsically tied to the density in overdense regions. Until recently, this density evolved in a manner very similar to the average density of the universe. It is only in more recent times that the structures in overdense regions have stabilized, and their dynamics have decoupled from the overall evolution of the universe.
	
	According to the proposed explanation, the universe's metric exhibits a profile across space that deviates significantly from perfect homogeneity. Notable variations exist between overdense and underdense regions, which are substantial enough to result in different dynamics for these regions compared to the universe as a whole. To assess the significance of these variations, it would be useful to quantify certain parameters from the two-regions model. For illustrative purposes (recognizing ongoing debate over precise values), we consider a measured Hubble constant of $H_{0(meas)} = 70\, (km/s)/Mpc$. Additionally, we assume that the density parameter for the apparent cosmological constant is currently $\Omega_{\Lambda 0} = 0.7$, while the density parameter for the apparent matter density is $\Omega_{M 0} = 0.3$.
	
	From the apparent dynamics observed in practice, as described by Eq. $(\ref{fqqx})$, we deduce
	\begin{eqnarray}
		\rho_\Lambda &=& \frac{3}{8\pi G} \Omega_{\Lambda 0} H^2_{0(meas)} = 6.44\times 10^{-27} \, kg/m^3 \,,
		\\
		\rho_{(meas)}(t_0) &=& \frac{3}{8\pi G} \Omega_{M 0} H^2_{0(meas)} = 2.76\times 10^{-27} \, kg/m^3 \,.
	\end{eqnarray}
	Recalling that $\rho (t_0) = \rho_{(meas)}(t_0)b_o^3(t_0)$, we deduce from Eq. $(\ref{da})$:
	\begin{equation}
		\rho_{(meas)}(t_0) = \frac{\rho(t_0)}{b_o^3(t_0)} = \frac{\rho_\Lambda}{\exp{\left(\frac{\rho_\Lambda}{\rho(t_0)}\right) - 1}}\,,
	\end{equation}
	from which we determine
	\begin{equation}
		\rho(t_0) = \frac{\rho_\Lambda}{\ln{\left(\frac{\rho_\Lambda}{\rho_{(meas)}(t_0)} + 1\right)}} = 5.36\times 10^{-27} \, kg/m^3 \,.
	\end{equation}
	Given that space is assumed to be globally flat, we can deduce the Hubble constant as it would be measured in a perfectly homogeneous universe from $\rho(t_0)$:
	\begin{equation}
		H_0 = \sqrt{\frac{3\pi G}{8}\rho(t_0)} = 53\, (km/s)/Mpc\,.
	\end{equation}
	This theoretical value of $H_0$	differs from the empirical measurements obtained through various observational methods, which have yielded a range of values close to $70\, (km/s)/Mpc$, some higher and some lower, depending on the technique used (see, for instance, \cite{Tully}). Notably, the methods used to infer the Hubble constant generally rely on observations of astrophysical objects such as galaxies, clusters, and supernovae. These objects tend to reside in overdense regions of the universe, which are not representative of the large-scale homogeneity assumed in cosmological models. The presence of these overdense structures introduces a systematic bias in local measurements of $H_0$. In the overdense regions where most of the data are collected, local gravitational effects can influence the inferred expansion rate, making it appear different from the true global value. This phenomenon is similar to the bias we are addressing here, as local measurements may not reflect the expansion rate of the universe as a whole, which is more accurately represented by the homogeneous cosmological model. Thus, while empirical estimates of the Hubble constant vary, they are often subject to local environmental influences that can distort the true global value, which in an idealized scenario would be closer to the $53\, (km/s)/Mpc$ derived from the assumption of a flat, homogeneous universe.
	
\begin{figure}
		\centering\includegraphics[width=10cm]{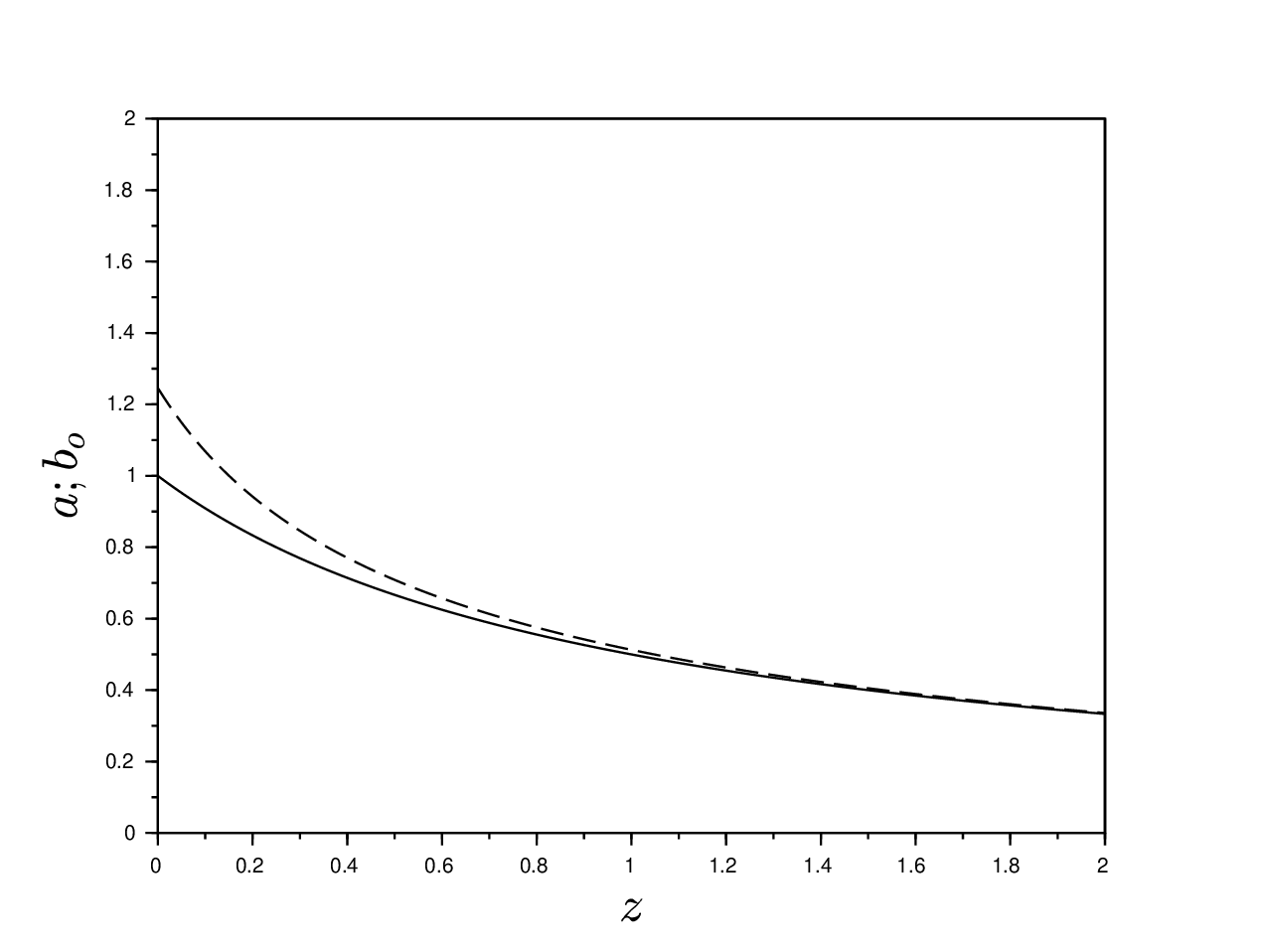}
		\caption{Evolution of $a$ (solid line) and $b_o$ (dashed line) in function of $z$.}\label{Fig1}
	\end{figure}

	With these values, we can determine the evolution of the scale factor $b_o$ in overdense regions over time and compare it with the scale factor $a$ of the universe. The results are illustrated in Figure $\ref{Fig1}$. As anticipated, at high redshift, $b_o$ closely matches $a$, indicating that the scale factor for the overdense regions closely matches the universal scale factor in the early universe. This convergence aligns with the idea that inhomogeneities were relatively small in the early universe, and the universe as a whole was more homogeneous. However, as the universe evolves toward lower redshift (closer to the present), the dashed line representing $b_o$ begins to diverge from the solid line representing $a$. This divergence highlights that the scale factor in overdense regions grows faster than the global scale factor of the universe. Currently, overdense regions have expanded such that the local scale factor is approximately 1.25 times larger than the scale factor of the universe.
	
\begin{figure}
		\centering\includegraphics[width=10cm]{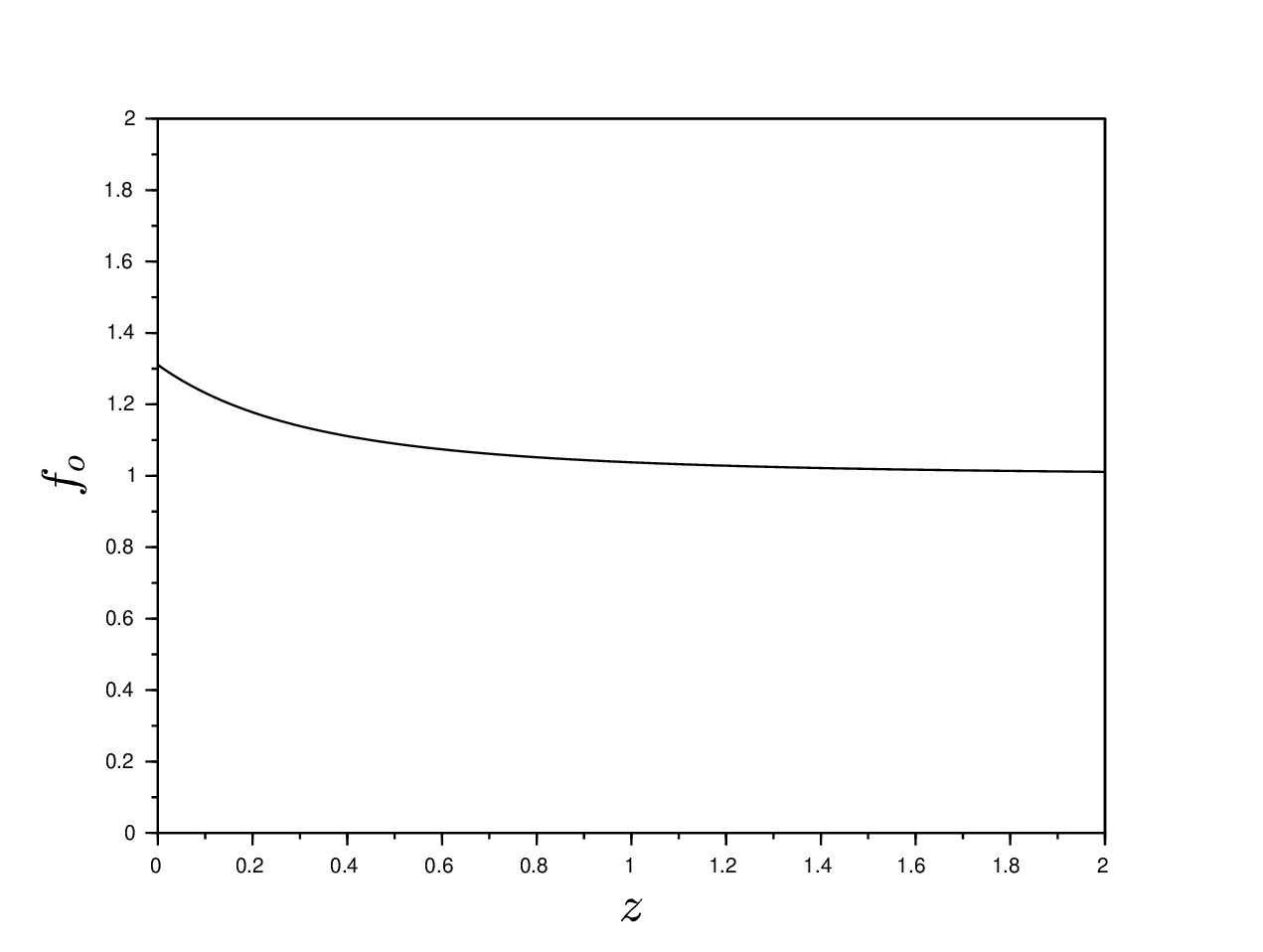}
		\caption{Evolution of $f_o$ in function of $z$.}\label{Fig2}
	\end{figure}

	In Figure $\ref{Fig2}$ we illustrate the evolution of $f_o$ as a function of redshift, highlighting how proper time in overdense regions compares to that of the universe. As expected, at high redshift, where inhomogeneities are minimal, $f_o$ approaches $1$. This indicates that, in the early universe, the rate at which time progresses in overdense regions is nearly the same as in the global, homogeneous universe. This matches the expectation that in the early universe, inhomogeneities were minimal, and both overdense regions and the universe as a whole evolved similarly. However, at lower redshift, it shows a significant deviation from this value. This deviation reflects the fact that overdense regions have a distinct local dynamics, leading to differences in how time is experienced within these regions compared to the larger universe. Currently, the rate at which proper time evolves in overdense regions is approximately 1.3 times greater than in the FLRW metric.
	
    This divergence in the scale factor and the faster evolution of proper time in overdense regions, as shown in Figures 1 and 2, point to the critical role that these local dynamics play in shaping the observations we interpret as cosmic acceleration. In the following discussion, we examine the broader implications of these local effects on our understanding of cosmic phenomena.
	
	
	If the observed expansion history is indeed influenced by the local dynamics of overdense regions, then this should manifest in measurable discrepancies between different cosmological probes. Since most measurements are taken from astrophysical objects residing in these regions, our inferred cosmological parameters, such as the Hubble constant or the equation of state of dark energy, may not accurately reflect the universe’s true large-scale behaviour. This offers a possible explanation for several long-standing observational tensions in cosmology. More exactly, we can offer a qualitative explanation for the tension observed between the SNIa and BAO probes, on one hand, and the Cosmic Microwave Background (CMB) probe, on the other. All three cosmic probes yield similar results, indicating an apparent acceleration in the expansion of the universe. However, the results obtained from the CMB probe show slight discrepancies when compared to those from the other two probes (see \cite{Riess2}).
	
	While the universe may exhibit a complex geometric structure, for practical purposes, we assume it can be globally characterized by the FLRW metric, which contains a single time-dependent parameter. The objective of our measurements is to optimally fit this FLRW metric to the actual data. The fitting process may vary depending on the cosmic probe used. To understand the intriguing tension problem, it is essential to have a clear understanding of the fitting processes associated with each cosmic probe. We must also remember that, according to the proposed explanation in this article, space is not perfectly homogeneous: the metric may exhibit significant spatial variations. For simplicity, we have considered two regions in this study, each characterized by a homogeneous metric. However, in reality, we should expect variations within those subregions as well, and not all overdense regions will present identical perturbations relative to the FLRW metric.
		
	The SNIa and BAO cosmic probes correspond to fitting processes that are conceptually similar. Measurements are conducted on large samples of events spanning different epochs and regions of the universe, albeit primarily within overdense regions. Each individual result is influenced by the local perturbations relative to the FLRW metric, but the fitting process averages these effects due to the size of the sample used. Ultimately, the resulting evolution can be interpreted as an average behaviour. As demonstrated, these two cosmic probes measure the average dynamics of the overdense regions.
	
	Conversely, the CMB cosmic probe employs a distinctly different fitting process. To begin, let us briefly outline the principle behind this probe. The cosmic microwave background is the remnant electromagnetic radiation from the early universe, permeating all of space. While the CMB is nearly isotropic, it exhibits slight anisotropies. When these anisotropies are analysed in terms of their power spectrum, specific peaks emerge. These peaks serve as signatures of various physical phenomena, such as the curvature of the universe, baryon density, and dark matter density. By analysing these parameters, we can indirectly infer the density of dark energy.
	
	The CMB probe is an indirect measure of the universe's dynamics. Unlike the SNIa probe, which directly measures the scale factor of the universe, the CMB probe infers the overall dynamics through a set of other measurements that reflect the conditions of the early universe. This approach is not problematic conceptually, as long as the underlying physics is well understood and accurately described. The equations governing the cosmological model that predicts CMB anisotropies form a coherent framework. By knowing certain cosmological parameters, we can subsequently deduce others, including those related to dark energy.
	
	However, the CMB probe has its limitations. Unlike the SNIa and BAO probes, which directly measure the scale factor at different points in time, the CMB does not provide direct measurements of the scale factor over cosmic history. This means that, while the CMB offers valuable insights into the early universe, it does not provide the same temporal resolution of the expansion history. Instead, the CMB probe assumes that the evolution of the scale factor follows the Friedmann equations, which describe the dynamics of a homogeneous and isotropic universe. The model that we fit to the real universe is constrained by this assumption and is typically represented by the $\Lambda$CDM model, with the cosmological constant $\Lambda$ being one of its key parameters. The values of other parameters, such as curvature, baryon density, and dark matter density, are derived from CMB measurements, and these in turn help estimate the properties of dark energy. While this assumption may appear limiting, the consistency of results from the SNIa and BAO probes lends significant support to the validity of this approach.

	Lastly, the CMB probe, as it can be applied, has a significant limitation. Unlike the SNIa and BAO probes, the CMB probe yields only a single measurement to fit the FLRW metric. This measurement is influenced by how the anisotropies are observed from our specific location, and it is inevitably affected by the metric perturbations present in the vicinity of Earth. Given that the Earth is situated in an overdense region, we should expect the results obtained from the CMB probe to reflect similar biases as those of the SNIa and BAO probes. However, while our simplified model assumes homogeneous overdense regions, in reality, there are likely perturbations within those regions. This suggests that the metric at the Earth's location may not perfectly match the average metric characterizing overdense regions. Since we have only one measurement to fit the FLRW metric, these perturbations are not averaged out, constraining the FLRW metric to correspond solely to the conditions at the Earth. In other words, this theory posits that the tension problem may stem from the discrepancy between the average metric in overdense regions and the actual metric in the Earth's vicinity. While this offers a plausible explanation, it is not a definitive proof, and further investigation is necessary to confirm or refute this hypothesis. However, such an exploration falls beyond the scope of this article.
		
	Beyond the discrepancies in inferred cosmic expansion rates, another observational puzzle challenges our understanding of the universe’s evolution: the unexpectedly early formation of galaxies. If the two-regions model accurately describes how local expansion rates influence observational measurements, it could also have implications for our interpretation of high-redshift objects. Recent observations by the James Webb Space Telescope (JWST) have detected galaxies that appear to have formed just a few hundred million years after the Big Bang, significantly earlier than conventional models expected, see \cite{Robertson}. These galaxies have been found at exceptionally high redshifts, with values as high as $z \sim 10-13$, indicating their presence within 300-400 million years after the universe's origin. However, it is important to recognize that these redshift values correspond to measured rather than actual values. In the very early universe, inhomogeneities were relatively small, meaning that the local scale factor $b_o$ closely approximates the universal scale factor $a$, as illustrated in Figure $\ref{Fig1}$. According to Eq. $(\ref{jj})$, the measured scale factor can then be expressed as $a_{meas} = a(t)/b_o(t_0)$, where $b_o(t_0)$, the scale factor in overdense regions at the present time, was estimated to be approximately $1.25$. This implies that the measured redshift for these early galaxies are overestimated by a factor of around 1.25. In other words, the redshift we observe today is inflated due to this discrepancy between local and universal scale factors, meaning that these galaxies might have formed later than the measured redshift suggests, thus easing the tension with current galaxy formation theories. This correction could provide a partial resolution to the unexpected rapidity of galaxy formation, aligning the observations with more conventional timelines while still accommodating the JWST's discoveries.
	
	If the two-regions model can explain both the apparent cosmic acceleration and the unexpectedly early formation of galaxies, an important next step is to find ways to test its predictions. A significant challenge in this endeavour is that all measurement processes involve both a source of information and an observer, and at least one of these is typically situated in an overdense region. This positioning can influence the results obtained from such measurements. Since current cosmological measurements predominantly sample overdense regions, validating the model would require alternative observational strategies. In particular, probing underdense regions, where local expansion may differ significantly, could provide crucial insights into whether the observed acceleration is truly a global effect or a consequence of measurement bias.
	
	It is important to note that underdense regions are not completely void of matter, they are capable of emitting information as well. However, phenomena such as Type Ia Supernovae (SNIa) are likely to be rare in these areas, and cosmic probes like the BAO and CMB are not well-suited for probing underdense regions. Therefore, the development of new methodologies for studying these regions is essential.
	
	One potential approach could involve a comparative analysis. If we can ascertain the proximity of an underdense region to an overdense one, we could analyse their respective redshifts. This comparison could help determine whether the redshift values are similar, as predicted by traditional cosmological models, or different, aligning with the expectations of the two-regions model. Such a comparative method could provide valuable insights into the dynamics of underdense regions and their relation to overdense regions. Unfortunately, practical difficulties remains, as cosmic voids are challenging to observe directly.
	
	Another potential validation approach involves the BAO cosmic probe. As previously discussed, the angular separation inferred from BAO measurements is derived from a mapping process that relies on redshift measurements. Consequently, any bias affecting redshift determinations could propagate into the inferred angular separation, potentially distorting the results. To test for such a bias, we propose a direct measurement of angular separations: instead of relying on redshift-derived mappings, we could consider all galaxies at the same observed redshift and measure the angular separation between them directly. By determining the BAO peak separation in this way, we would obtain an independent value, unaffected by redshift-based reconstruction. If this directly measured angular separation differs from the one inferred through standard mapping techniques, it would provide strong evidence for the existence of an observational bias affecting redshift-based measurements. However, implementing this method would require addressing challenges such as measurement precision and the need for a sufficiently large and well-distributed sample of galaxies to reliably extract the BAO signal.
	
	
	\section{Conclusion}
	
	This article has explored the possibility that the observed accelerated expansion of the universe, commonly attributed to dark energy, may instead be an artefact of observational biases caused by the inhomogeneous structure of the universe. By refining the two-regions model, which accounts for the different dynamics in overdense and underdense regions, we have shown that measurements from cosmic probes such as Type Ia supernovae and baryon acoustic oscillations may predominantly reflect the local dynamics of overdense regions rather than the global expansion of the universe.
	
	Our analysis reveals that the observed dynamics of the universe corresponds to a total density that evolves similarly to what one would expect in overdense regions. Since the majority of cosmic measurements are taken from objects situated in overdense regions, one might wonder whether this apparent coincidence could provide a vital clue to the true nature of dark energy. This potential bias, inherent in the measurement process, suggests that what we interpret as global acceleration might in fact be a reflection of the local, overdense dynamics, calling into question the necessity of invoking dark energy to explain these observations.
	
	While this model provides a compelling alternative to dark energy, it also provides potential solutions to several related challenges. The model provides a natural resolution to the coincidence problem. The near-equality of matter and dark energy densities at the present epoch has long been considered an unexplained fine-tuning issue. However, our findings indicate that this near-equality emerges naturally from the stabilization of densities within overdense regions. As these regions become gravitationally bound, their density evolution mimics the behaviour of a cosmological constant, effectively generating the observed acceleration without requiring an additional dark energy component. Furthermore, our approach sheds new light on ongoing cosmological tensions, and also offers a potential resolution to the unexpectedly early formation of galaxies observed by the James Webb Space Telescope (JWST). By accounting for the local scale factor, which inflates measured redshifts by a factor of about 1.25, the actual formation of these galaxies could have occurred later than initially inferred, bringing the observations closer to standard cosmological predictions.
	
	Looking forward, further observational tests are needed to validate this model. A key challenge lies in probing cosmic expansion in underdense regions, where local expansion may differ significantly from that in overdense regions. Directly measuring the angular separation of BAO peaks, rather than inferring it from redshift-based mappings, could provide an independent test of the proposed observational bias. Additionally, comparative analyses of redshifts in neighbouring overdense and underdense regions could offer direct empirical evidence for the differing local expansion rates predicted by this model.
	
	Ultimately, this study highlights the necessity of considering inhomogeneities when interpreting cosmological observations. As precision measurements continue to refine our understanding of the universe’s structure and expansion, the two-regions model provides a compelling alternative to dark energy, potentially reshaping our fundamental understanding of cosmic evolution. If confirmed, this perspective would not only resolve several long-standing cosmological puzzles but also shift the focus of future research towards a more nuanced exploration of the role of structure formation in shaping our observations of the universe.
	


	
\end{document}